\documentclass[twocolumn,aps,prl,epsf,graphics,psfig]{revtex4}
\usepackage{graphicx}
\begin{document}
\title{Electronic States of Magnetic Quantum Dots}

\author{Ramin M. Abolfath$^{1,2,3}$}
\author{Pawel Hawrylak$^2$}
\author{Igor \v{Z}uti\'c$^1$}
\affiliation{
$^1$ Department of Physics, State University of New York at Buffalo,
Buffalo, New York 14260, USA \\
$^2$Institute for Microstructural Sciences,
National Research Council of Canada, Ottawa, K1A 0R6, Canada \\
$^3$Department of Radiation Oncology, University of Texas Southwestern
Medical Center, Dallas, Texas 75390, USA
}


\begin{abstract}
We study quantum states of electrons in magnetically doped 
quantum dots as a function of exchange coupling between 
electron and impurity spins, the strength of Coulomb interaction, 
confining potential, and the number of electrons.
The magnetic phase diagram of quantum dots,  
doped with a large number of magnetic Mn impurities, can be 
described by the energy gap in the spectrum of electrons
and the mean field electron-Mn exchange coupling.
A competition between these two parameters leads to a 
transition between spin-unpolarized and spin-polarized states,
in the absence of applied magnetic field. Tuning the energy gap 
by electrostatic control of nonparabolicity of
the confining potential can enable control of 
magnetization even at the fixed number of electrons.
We illustrate our findings by directly comparing Mn-doped 
quantum dots with parabolic and Gaussian confining potential. 
\end{abstract}
\maketitle

Recent experimental~\cite{Mackowski2004:APL,Gould2006:PRL,%
Besombes2004:PRL,Leger2006:PRL,Chakrabarti2005:NL,Wojnar2007:PRB,%
Archer2007:NL,Liu2006:JACS,Zhen2007:CPL}
and theoretical~\cite{Hawrylak1991:PRB,%
Bhattacharjee1997:PRB,Fernandez-Rossier2004:PRL,%
Govorov2005:PRB,Qu2005:PRL,Abolfath2007:PRL}
studies reveal that semiconductor quantum dots
 (QDs)~\cite{Jacak1998:Book,%
Reimann2002:RMP}
are desirable for tailoring magnetism which persists at much high 
temperatures than in their bulk counterparts~\cite{Dietl1997:PRB}.  
These nanoscale realizations of dilute magnetic 
semiconductors (DMS)~\cite{Furdyna1988:JAP} 
may be suitable for a versatile control of spin and magnetism
with potential applications in information storage technology
and spintronics~\cite{Zutic2004:RMP}.
An important property of the carrier-mediated ferromagnetism 
in bulk-like DMS is the optical and electrical control of 
the magnetic ordering by changing the number of 
carriers~\cite{Koshihara1997:PRL,Ohno2000:N,Boukari2002:PRL}.
However, because of the strong interplay between quantum confinement 
and the many-body electron-electron (e-e) Coulomb interactions in QDs,
it may be possible to achieve and control the magnetic ordering of
 carrier 
spin and magnetic impurities, with a fixed number 
of carriers~\cite{Abolfath2007:PRL}.  

In this paper, we present a theoretical analysis
to describe the quantum states of the electrons confined
in II-VI quantum dots doped with Mn.
In (II,Mn)VI materials, Mn does not change
the number of electrons ($N$) because it is isoelectronic 
with group-II elements.
In QDs, however, additional carriers
are controlled by either chemical doping or by an external
 electrostatic
potential applied to the metallic gates. 
Even for a small number of interacting electrons, the study of QDs 
doped with large numbers of Mn ($> 10$) becomes computationally 
inaccessible to  exact diagaonalization 
techniques~\cite{Qu2005:PRL,Abolfath2007:PRL}.
For this reason we employ here a mean field theory of magnetic QDs 
in which Mn acts like an external magnetic field and the weak coupling 
between electrons and magnetic impurities 
can be described as spin-spin exchange
 interaction~\cite{Abolfath2007:PRL}.
Because electrons are confined in all three dimensions in
 nanometer-sized QDs,
we anticipate strong electron-electron (e-e) Coulomb interaction.
We use a finite temperature local spin density approximation 
(LSDA)~\cite{Dharma-wardana1995:Book} to incorporate e-e Coulomb
 interaction
in this work.

The wave function of electrons in QDs can be expanded in terms of its 
planar $\psi_{i\sigma}(\vec{\rho})$ with $\vec{\rho}\equiv (x,y)$, 
and subband wave function $\xi(z)$.
The effective two-dimensional (2D) Kohn-Sham (KS) equations
$H \psi_{i\sigma}(\vec{\rho}) 
= \epsilon_{i\sigma}\psi_{i\sigma}(\vec{\rho})$,
in LSDA can be obtained by integrating $\xi(z)$~\cite{Xia1992:PRB}, 
assuming that the first subband is filled.
Here $\epsilon_{i\sigma}$ are the KS eigenenergies, 
$\sigma=+1$ and $-1$ for spin up ($\uparrow$) and down ($\downarrow$), 
and $H$ is the Kohn-Sham Hamiltonian which can be expressed as
\begin{eqnarray}
H = \frac{-\hbar^2}{2m^*} \nabla_\rho^2
+ V_{QD} + \gamma V_{H} + \gamma V^\sigma_{XC}
- \frac{\sigma}{2} h_{sd}(\vec{\rho}).
\label{Heff}
\end{eqnarray}
where $\hbar$ is the Planck constant, $m^*$ is the electron effective
 mass, 
and $V_{QD} = V_0 \exp(-\rho^2/\delta^2)$ is the 2D Gaussian 
confining potential 
of the quantum dot ( 1D parabolic confinement is chosen
along the $z$-axis). 
In Eq.(\ref{Heff}), $V_H$ and $V^\sigma_{XC}$ are Hartree and 
spin dependent exchange-correlation
 potentials~\cite{Dharma-wardana1995:Book},
while $\gamma$ accounts for reduction of Coulomb strength due to
 screening
effects of the gate electrodes~\cite{Bruce2000:PRB}
and
\begin{eqnarray}
h_{sd}(\vec{\rho}) =  J_{em} \int dz |\xi(z)|^2 
B_M\left( \frac{M b(\vec{\rho}, z)}{k_BT}\right).
\label{hsd}
\end{eqnarray}
$J_{em} = J_{sd} n_m M$ is the e-Mn exchange coupling, $J_{sd}$
is the exchange coupling between electron and single Mn, $n_m$ is the
spatial-averaged density of Mn, and $M=5/2$ is the spin of Mn.
$B_M(x)$ is the Brillouin function~\cite{Ashcroft:1976} in which
the argument $x$ contains 
$k_B$ the Boltzmann constant, $T$ the absolute temperature, and 
\begin{eqnarray}
b(\vec{\rho}, z) =- J^{AF}_{\rm eff} \langle M_z(\vec{\rho}, z) \rangle
+ \frac{J_{sd}}{2} [n_\uparrow(\vec{\rho}, z) -
 n_\downarrow(\vec{\rho}, z)],
\label{beff}
\end{eqnarray}
the effective field seen by the Mn~\cite{Abolfath2007:PRL}. 
The first term in $b(\vec{\rho}, z)$ describes the mean field
of the direct Mn-Mn antiferromagnetic 
coupling~\cite{Fernandez-Rossier2004:PRL}, with 
\begin{eqnarray}
\langle M_z(\vec{\rho}, z)\rangle = M 
B_M\left(\frac{M b(\vec{\rho}, z)}{k_BT}\right),
\label{Meff}
\end{eqnarray}
and 
\begin{eqnarray}
n_\sigma(\vec{\rho}, z) = \sum_i |\psi_{i\sigma}(\vec{\rho})|^2
 |\xi(z)|^2 
f(\epsilon_{i\sigma}),
\label{density}
\end{eqnarray}
is the spin-resolved electron density, 
$f(\epsilon)=1/\{\exp[(\epsilon-\mu)/k_BT]+1\}$ is the Fermi function,
and $\mu$ is the temperature-dependent chemical potential, and is
the self-consistent solution of $N = \sum_{i,\sigma}
 f(\epsilon_{i\sigma})$
which ensures that $N=N_\uparrow + N_\downarrow$ is an integer number.

To obtain solutions of Eqs.~(\ref{Heff})-(\ref{density}), 
we consider (Cd,Mn)Te QD with $V_0=-128$ meV, and $\delta=15.9$ nm, 
a perpendicular width of 1 nm, $J_{sd}=0.015$ eV nm$^3$, 
$n_m=0,~0.025,~0.1$ nm$^{-3}$, corresponding to 
the approximate number of Mn, $N_{\rm Mn}=0,~45,~180$,
$J_{em} =0,~0.94,~3.75$ meV,
and $J^{AF}_{\rm eff}=0,~0.005,~0.02$ meV, respectively.
For CdTe, considered here,  $a^*_B=5.29$ nm, and $Ry^*=12.8$ meV are
 the 
effective Bohr radius and Rydberg energy, while the additional
material parameters are $m^*=0.106$, and
$\epsilon=10.6$~\cite{Abolfath2007:PRL}.

\begin{figure}
\begin{center}\vspace{1cm}
\includegraphics[width=0.8\linewidth]{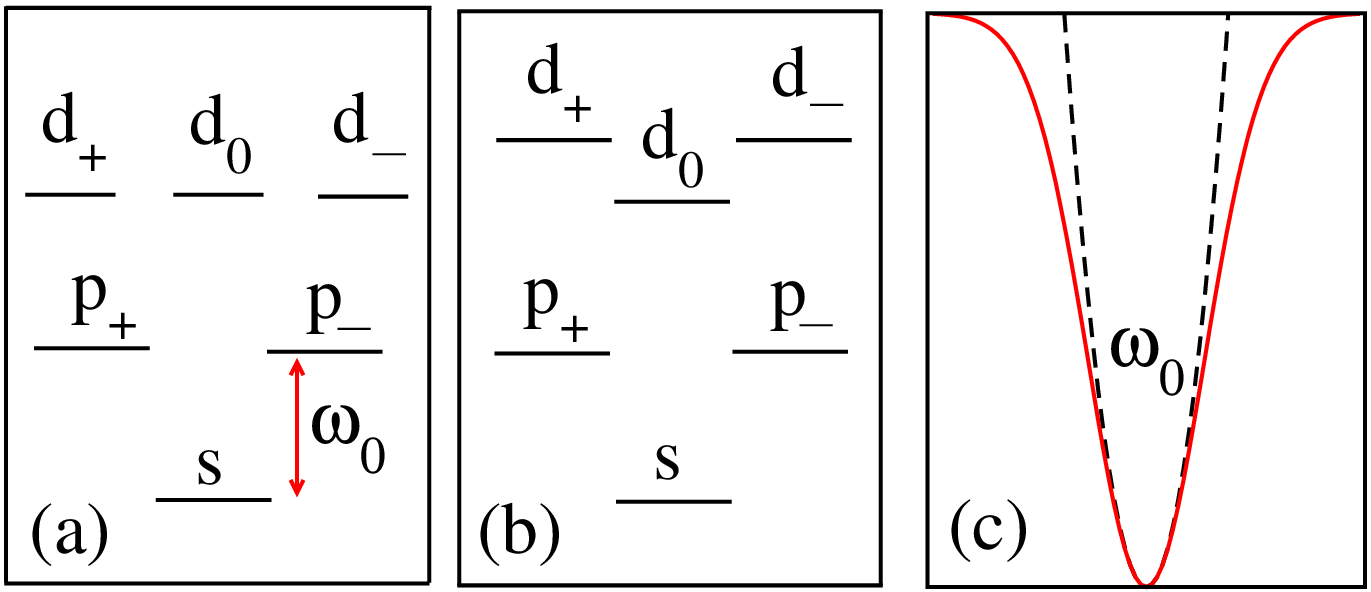}
\caption{
The schematic single particle levels of a 2D parabolic (a) and
2D Gaussian (b) confining potentials corresponding to dashed and bold 
lines in (c). 
}
\label{ffig0}
\end{center}
\end{figure}

The shell structure of the 2D parabolic (a) and 2D
Gaussian (b) potentials are shown in Fig.~\ref{ffig0}.
The corresponding confining potentials are shown in 
Fig.~\ref{ffig0}(c) where the dashed and bold lines represent
the 2D parabolic and 2D Gaussian potentials.
The distinction between parabolic and Gaussian potentials can 
be experimentally resolved by photo-emission spectroscopy or 
transport measurements of single particle states in external 
magnetic field \cite{Jacak1998:Book,Reimann2002:RMP}.
The energy gap between $s$-, $p$-, and $d$-orbitals 
is characterized by $\omega_0$. 
For a 2D Gaussian potential, 
the characteristic energy associated with the perpendicular 
confinement, $\omega_0$, is calculated by expanding $V_{QD}$ 
in the vicinity of the minimum which yields 
$V_{QD}=V_0 + m^* \omega_0^2 \rho^2/2 + \dots$, 
with the strength $\omega_0=\sqrt{2|V_0|/m^*} /\delta$.
Because of the circular symmetry of the Gaussian potential, the
 $z$-component
of the angular momentum $l_z$ is a good quantum number.
As a result of the geometrical symmetry of the confining potential, 
the states in $p$-shell with $l_z=\pm 1$ are degenerate. 
In contrast to the 2D parabolic
 potential~\cite{Jacak1998:Book,Reimann2002:RMP},
where the dynamical symmetry of the confining potential implies  
occurrence of accidental degeneracies of the single particle levels 
with different $|l_z|$,
the single particle levels in $d$-shell are not completely degenerate.
Degenerate levels $d_+$ and $d_-$ are separated by an energy gap
($1.5$ meV) from the $d_0$-level, 
where the indices $\pm,0$ refer to angular momentum $l_z = \pm 1,0$.
We note that the ordering of the single particle 
$s$-, $p$-, and $d$-eigenenergies in the Gaussian potential are similar
 to the ordering of the first few single particle eigenenergies 
in a square potential.

\begin{figure}
\begin{center}\vspace{1cm}
\includegraphics[width=0.9\linewidth]{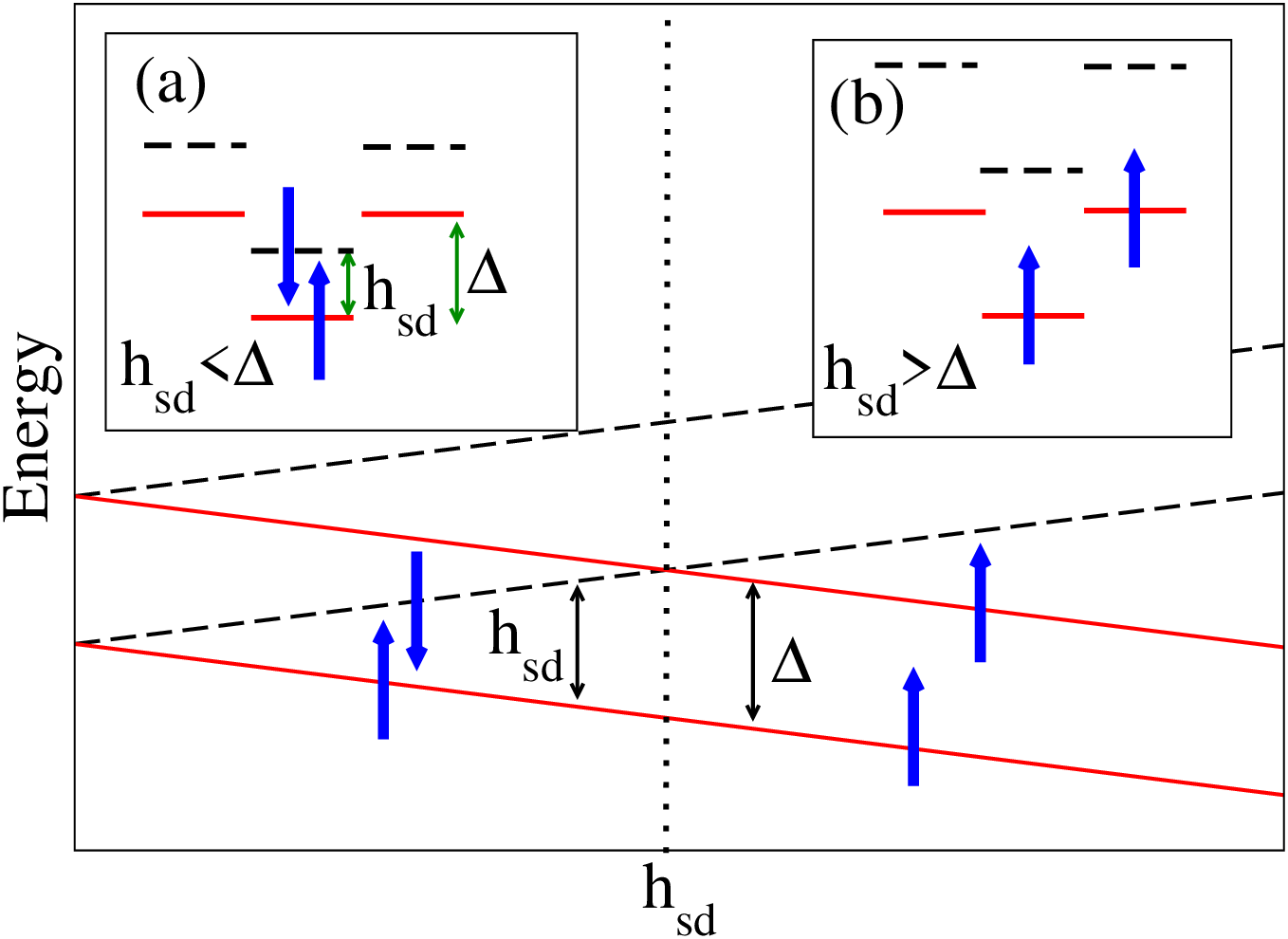}
\caption{
Single particle levels of the $d$-shell as a 
function of effective Zeeman energy $h_{sd}$.
The insets (a) and (b) schematically show the ground state of $N=8$
depending on the relative magnitude of $h_{sd}$ and the single
particle energy gap $\Delta$.
At $\Delta=h_{sd}$ (shown by dashed line) a transition from 
antiferromagnetic to ferromagnetic state is predicted.
Filled and dashed lines represent the single particle eigenenergies 
of the electrons with spin up and down.
}
\label{ffig1}
\end{center}
\end{figure}

In the absence of direct antiferromagnetic coupling between Mn,
and e-e Coulomb interaction,
an identification of the magnetic states of the QDs based on 
the present mean field theory can be expressed in terms of two energy
 scales:
the single particle energy gap $\Delta$, and the effective e-Mn
 exchange 
coupling $J_{em}$.
The former is the energy difference between the highest occupied
 molecular 
orbital (HOMO) and the lowest unoccupied molecular orbital (LUMO).
It is a function of the number and spin of electrons, 
and the geometrical and dynamical symmetries of the QD confining
 potential.
Therefore $\Delta$, which depends on the quantum confinement,
can be tuned and controlled by the electric 
voltage applied to the metallic gates.
On the other hand, $J_{em}$, 
the strength of the effective Zeeman energy, 
$h_{sd} \equiv \langle h_{sd}(\vec{\rho}) \rangle$,
which polarizes spin of electrons, 
is independent of the quantum confinement.
Thus the magnetic phase diagram of the QDs can be effectively described
 in 
terms of $\Delta$, and $J_{em}$. In the absence of magnetic field we
 can 
then distinguish between spin-polarized and spin-unpolarized states in
 QDs, 
conventionally also referred to as the ferromagnetic
(FM) and antiferromagnetic (AFM) 
states~\cite{Fernandez-Rossier2004:PRL,Govorov2005:PRB,Qu2005:PRL}.
For example, the ground state of the electrons exhibits FM ordering 
if $J_{em} > \Delta$.
The dependence of the single particle eigenenergies of the $d$-shell 
with $N=8$ on $h_{sd}$ is shown in Fig.~\ref{ffig1}.
At $h_{sd}=0$ the valence electrons with spin up and down occupy single
 particle level $d_0$, 
following the Pauli exclusion principle, and the ground state of $N=8$ 
exhibits AFM ordering with $N_\uparrow = N_\downarrow =4$.
With increasing $h_{sd}$ the gap between spin up and spin down opens
 up.
At $h_{sd} = \Delta = 1.5$ meV, where $\Delta=E_{d_+(d_-)} - E_{d_0}$, 
the first spin flip occurs.
Because of well separated $p$- and $d$- orbitals the second spin flip 
occurs at much higher energies as $h_{sd} = \omega_0 \approx 7 J_{em}$.

\begin{figure}
\begin{center}\vspace{1cm}
\includegraphics[width=0.9\linewidth]{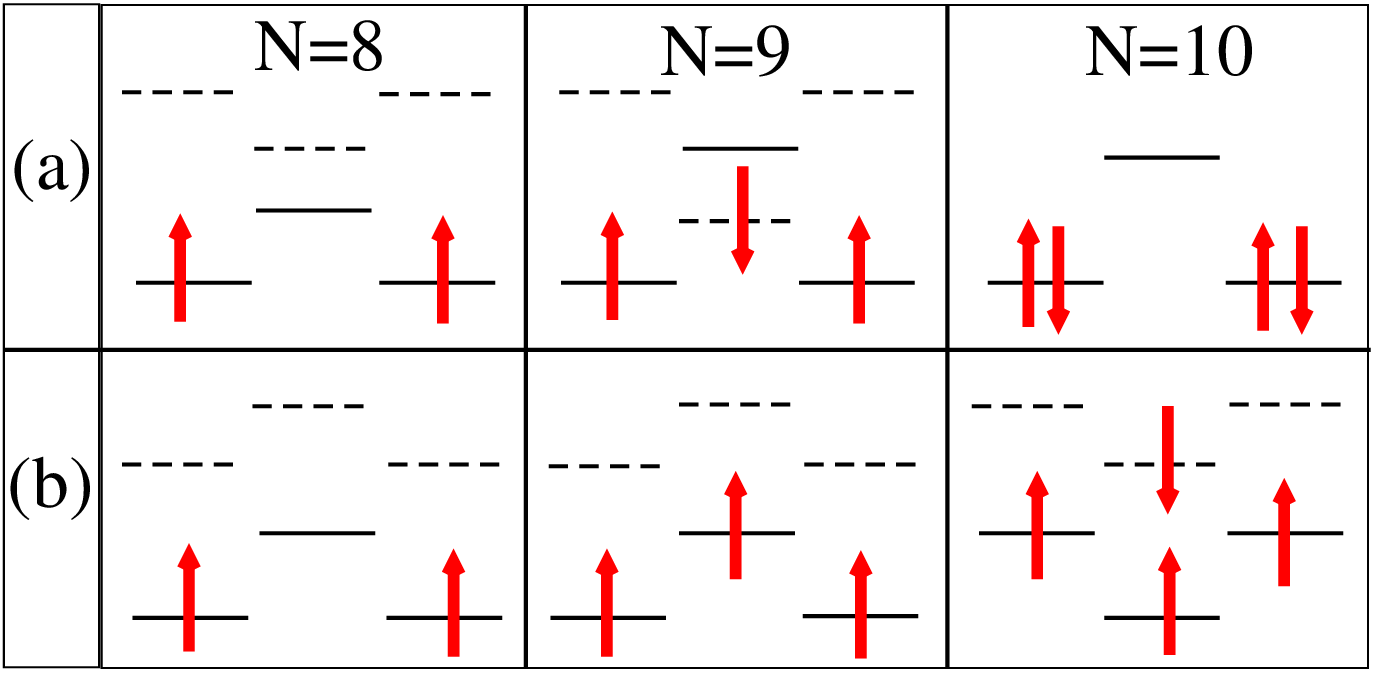}
\caption{
Spin up (bold lines), and spin down (dashed lines) show
Kohn-Sham eigenstates of $N=8,9,10$ electrons in $d$-shell in 2D
 Gaussian 
confining potential with $V_0=-128$ meV, $\delta=15.9$ nm,
$n_m=0$ (a), and $n_m=0.1$ nm$^{-3}$ (b). For clarity, 
$s$- and $p$-levels are not shown.
}
\label{ffig2}
\end{center}
\end{figure}

\begin{figure}
\begin{center}\vspace{1cm}
\includegraphics[width=0.8\linewidth]{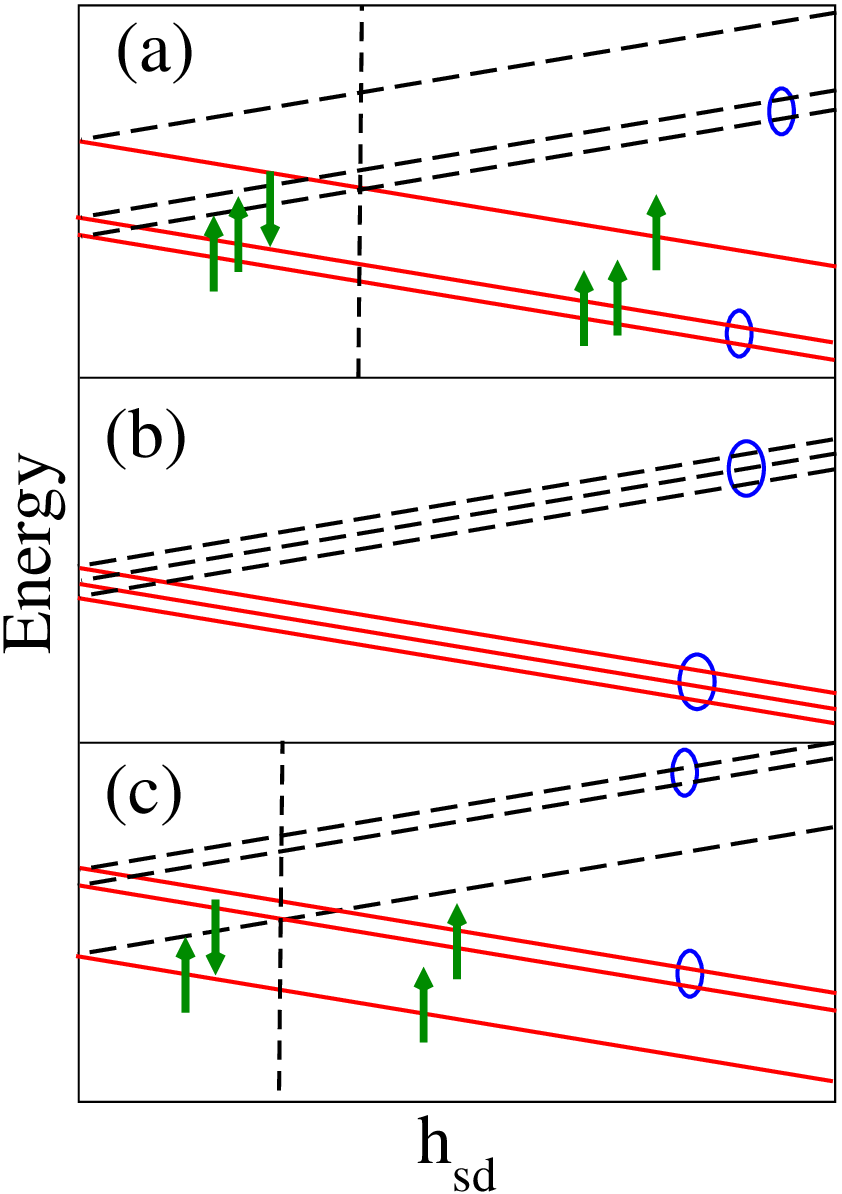}
\caption{
Energy of quantized states in 2D Gaussian confining potential  
with spin up (bold lines) and spin down (dashed lines) in $d$-shell
as a function of effective Zeeman energy $h_{sd}$ corresponding to
$\Delta^*<0$ (a), $\Delta^*=0$ (b), and $\Delta^*>0$ (c).
Energy levels marked with circles are degenerate and 
shifted with respect to each other for enhanced visibility. The
first spin flip occurs if $\Delta^*<0$, $N=9$ (a), and $\Delta^*>0$,
 $N=8$ (c).
No spin flip occurs if $\Delta^*=0$ (b).
}
\label{ffig20}
\end{center}
\end{figure}

In LSDA, electrons are replaced by independent
quasi-particles (electrons dressed by 
e-e interaction), where
KS eigenenergies are the energy levels of the quasi-particles.
In this picture, e-e interaction rearrange the structure 
of quantized levels, and  affects the shell structure of the QDs.
In this limit, the quasi-particle gap $\Delta^*$ 
(renormalized $\Delta$) is the energy 
difference between HOMO and LUMO of the KS eigenenergies.
It is an appropriate 
parameter, which can supersede the single particle gap $\Delta$
in identification of the magnetic states of the QDs.

The KS levels in the $d$-shell of the 2D Gaussian potential 
with $n_m=0$ and $N=8,9,10$ are shown in 
Fig.~\ref{ffig2}(a). 
Because of circular symmetry of the confining potential,
the energy levels of $d_+$ and $d_-$ are degenerate.
The energy of $d_0$ which was lower than $d_+$ and $d_-$ due
to the nonparabolicity of the Gaussian potential, increases by
e-e Coulomb interaction, such that it becomes the highest energy level
in $d$-shell, and $\Delta^*$ changes sign.
In contrast to non-interacting electrons, the ordering of the 
KS levels in $s$-, $p$-, and $d$-shells is closer to the one in
 circular 
potential than in the square potential.
In the absence of magnetic impurities, as it is shown in
 Fig.~\ref{ffig2}(a),
$N=8$ forms a half-filled shell with $s_z=1$, and
the electron polarization $P=2/8$, and
$N=9$ shows an open shell with $s_z=1/2$, and $P=1/9$.
Because of $d$-shell overturning, 
$N=10$ forms a closed shell with $s_z=0$, and $P=0/10$.
Here the energy difference between the KS eigenenergies of the
 electrons 
with spin up and down is zero, and $\Delta^* \approx 0.6$ meV
($< J_{em}$).
With increasing e-Mn coupling to $J_{em}= 3.75$ meV, corresponding to 
5\% doping, 
and within a range that spin polarization happens only in $d$-shell,  
we find no magnetic transition for $N=8$, since
it was already fully polarized due to the spin Hund's rule,
whereas for $N=9$ and $N=10$ we find transitions to $s=3/2$ 
and $s=1$ respectively. 
The evolution of quantized energies as a function of effective Zeeman
 energy
and $\Delta^*$ is shown in Fig.~\ref{ffig20}. 
For examples $\Delta^*<0$ corresponds to $d$-levels of the 2D
Gaussian potential with $\gamma=1$, whereas $\Delta^*=0$ (b), 
and $\Delta^*>0$ (c) correspond to $d$-levels in the 2D parabolic 
and 2D Gaussian potentials with $\gamma=0$ ($\Delta^* = \Delta$).

\begin{figure}
\begin{center} 
\includegraphics[width=0.9\linewidth]{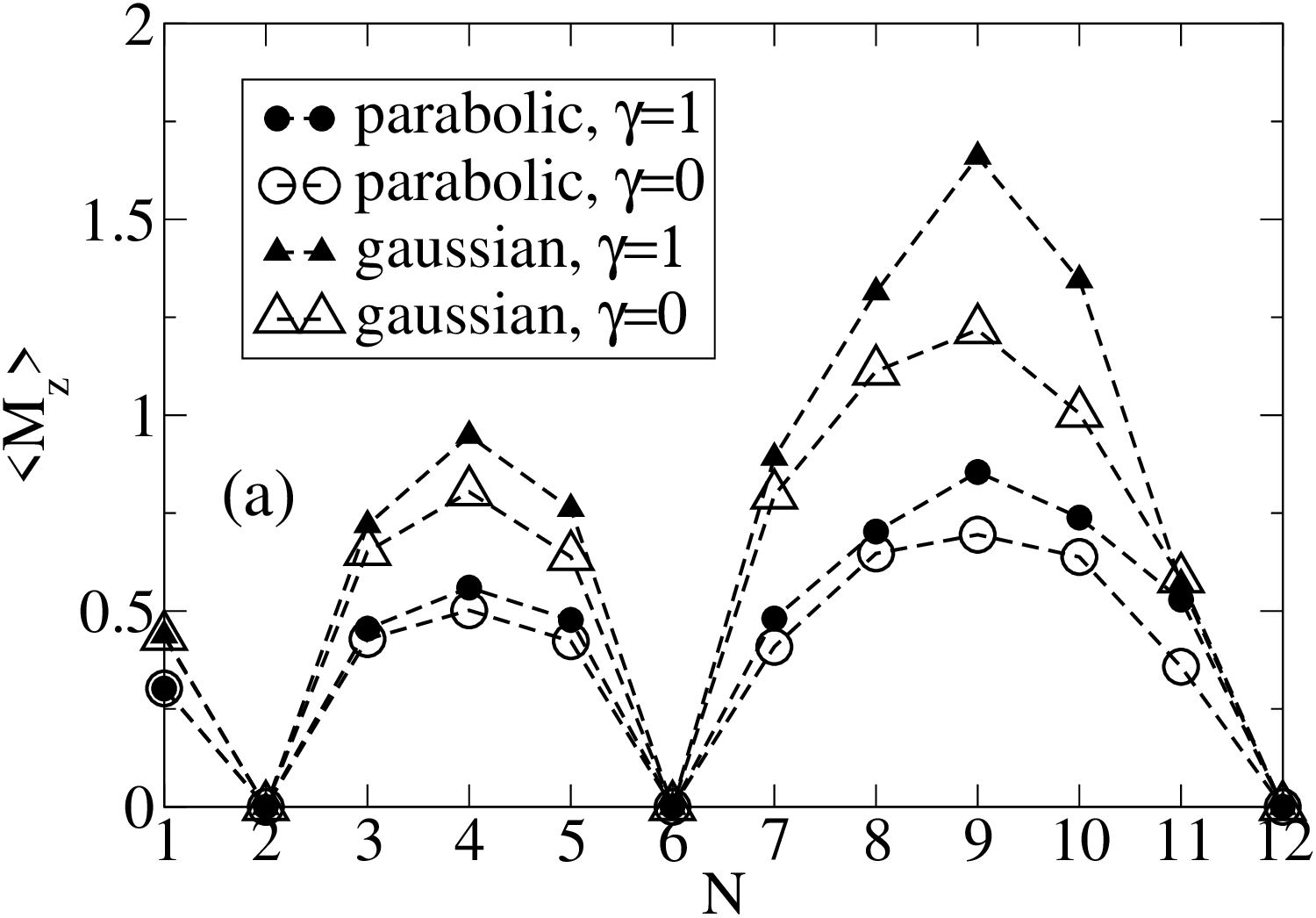}\vspace{1cm}
\includegraphics[width=0.9\linewidth]{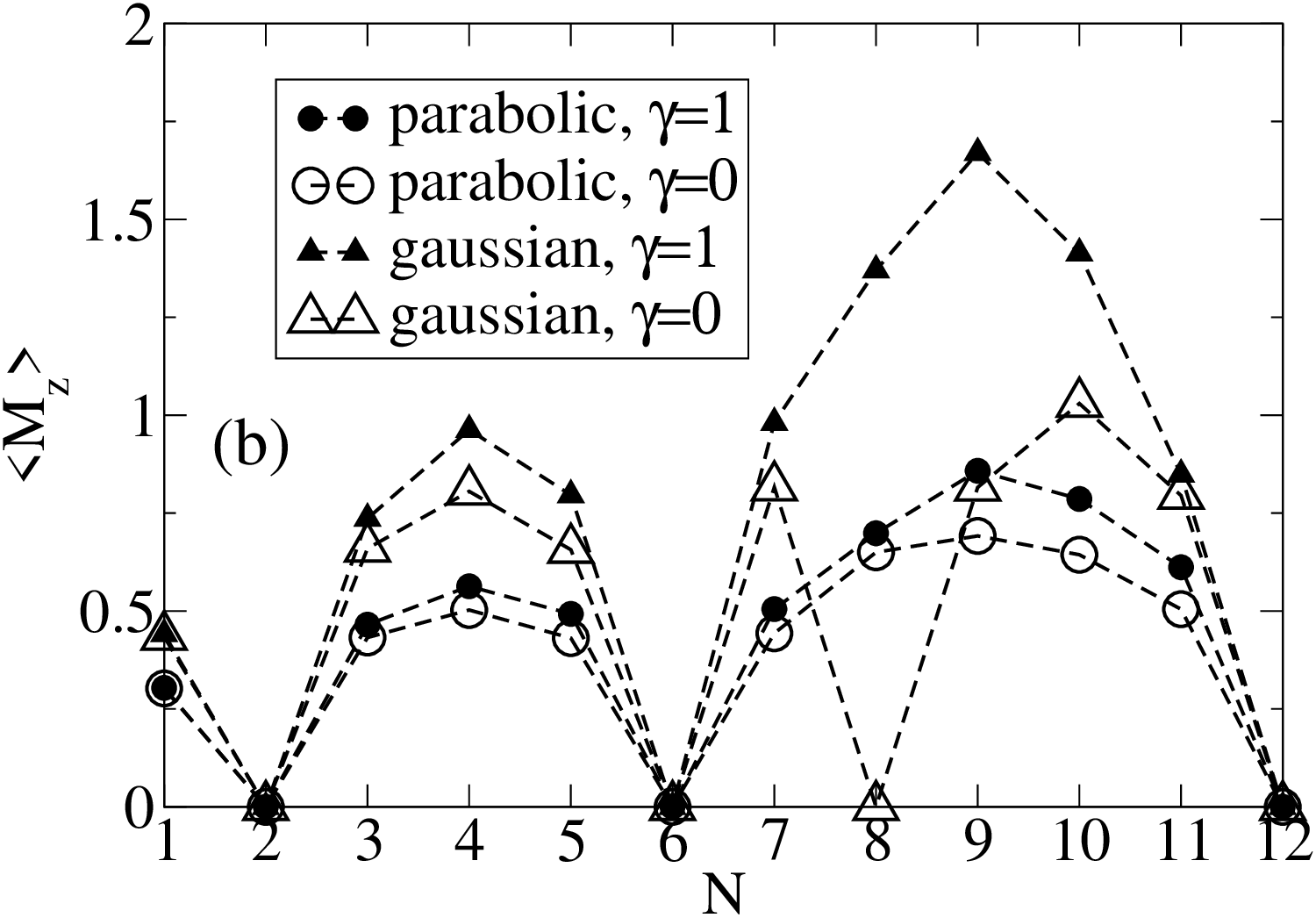}
\caption{
The averaged magnetization per unit area
$\langle M_z\rangle$ as a function of number of electrons
$N$ at $T=1K$ and Mn-density $n_m=0.1$ nm$^{-3}$ (a), 
and $n_m=0.025$ nm$^{-3}$ (b) for
non-interacting ($\gamma=0$, empty triangles) and 
interacting ($\gamma=1$, filled triangles) electrons.
At $N=8$, $n_m=0.025$ nm$^{-3}$, in 2D Gaussian confining potential
the ground state of the QD switches between ferromagnetic
and antiferromagnetic states by e-e Coulomb interaction. 
}
\label{ffig3}
\end{center}
\end{figure}

We next turn to calculate the spatially-averaged Mn-magnetization
per unit area $A$, $\langle M_z \rangle = \frac{1}{A} 
\int d^2\rho \langle M_z(\vec{\rho}) \rangle$
for the 2D parabolic and 2D Gaussian potentials 
as a function of $N$ for zero ($\gamma=0$), and full ($\gamma=1$)
e-e Coulomb interaction in Fig.~\ref{ffig3}.
This calculation illustrates how deviation 
from nonparabolicity of the confining
potential may lead to changes in the magnetic properties of QDs.
Many qualitative features indeed coincide for both of the confinements.
 
A comparison in  Fig.~\ref{ffig3}(a) for $n_m=0.1$ nm$^{-3}$, confirms 
that there is similarity in magnetization for both Gaussian and
 parabolic 
confinements.
Here $J_{em} > \Delta^*~(\Delta)$ for $\gamma=1~(0)$ and we consider
 all 
the electron numbers up to a full $d$-shell ($N\le 12$).
However, as we show in Fig.~\ref{ffig3}(b) a 
nonparabolic confinement can introduce additional magnetic transitions 
in quantum dots.
Here $n_m=0.025$ nm$^{-3}$ corresponds 
to a magnetic doping of 1.25 \%. 
At $N=8$ and $\gamma=0$ in the 2D Gaussian potential,
we find that $J_{em} < \Delta$ and a transition to the AFM state.
In contrast, at $\gamma=1$, we find $J_{em} > \Delta^*$, and thus the
 FM state
is stable.

Our findings illustrate the importance of Coulomb interactions
and quantum confinement for the magnetic ordering of carrier spin
and magnetic impurities in (II,Mn)VI quantum dots. We show that
with electrostatic control of nonparabolicity of the confining
 potential,
it is possible to control the magnetization of QDs even with a fixed 
number of electrons. While the choice of 2D Gaussian confinement is 
a convenient form to parameterize nonparabolic effects, other
 deviations
from the parabolic confinement could also influence the magnetic
 ordering 
in quantum dots \cite{condmat}.

This work is supported by the US ONR, NSF-ECCS CAREER, the NRC HPC
 project,
CIAR, the CCR at SUNY Buffalo, and the Center for Nanophase Materials
Sciences, sponsored at ORNL by the Division of Scientific User
 Facilities, 
US DOE.

       


\end{document}